\documentclass[twocolumn]{emulateapj}
\usepackage[latin2]{inputenc}

\renewcommand{\vec}[1]{\mbox{\boldmath$#1$}}
\newcommand{\bssC}{\textbf{\textsf{C}}}
\newcommand{\R}{\mathcal{R}}
\newcommand{\avg}[1]{\left\langle{#1}\right\rangle}

\newcommand{\br}{{\vec r}}

\newcommand{\hmpc}{\,$h^{-1}$\,Mpc}
\newcommand{\ihmpc}{\,$h$\,Mpc$^{-1}$}
\newcommand{\LCDM}{$\Lambda$CDM}
\newcommand{\pd}{$P_\delta$}
\newcommand{\pdglob}{$P_{\delta,\, {\rm global}\, \bar{\rho}}$}
\newcommand{\plog}{$P_{\log(1+\delta)}$}
\newcommand{\dg}{$\delta_{\rm Gauss}$}
\newcommand{\pg}{$P_{\rm Gauss}$}
\newcommand{\camb}{{\scshape camb}}
\newcommand{\Var}{{\rm Var}}
\newcommand{\eg}{e.g.,}

\chardef\til=`\~

\begin{document}

\title{Rejuvenating the matter power spectrum: restoring information
  with a logarithmic density mapping}

\author{Mark C.\ Neyrinck\altaffilmark{1}, Istv\'an Szapudi\altaffilmark{2,3,4} and Alexander S.\ Szalay\altaffilmark{1}}
\altaffiltext{1}{Department of Physics and Astronomy, The Johns Hopkins University, 3701 San Martin Drive, Baltimore, MD 21218, USA}
\altaffiltext{2}{Institute for Astronomy, University of Hawaii, 2680 Woodlawn Drive, Honolulu HI 96822, USA}
\altaffiltext{3}{Institute for Advanced Study, Collegium Budapest, Szentháromság u.\ 2., Budapest, H-1014, Hungary}
\altaffiltext{4}{Eötvös Loránd University, Dept.\ of Atomic Physics, 1117 Pázmány Péter sétány 1/A}

\begin{abstract}
We find that nonlinearities in the dark-matter power spectrum are
dramatically smaller if the density field first undergoes a
logarithmic mapping.  In the Millennium simulation, this procedure
gives a power spectrum with a shape hardly departing from the linear
power spectrum for $k\lesssim 1$\ihmpc\ at all redshifts.  Also, this
procedure unveils pristine Fisher information on a range of scales
reaching a factor of 2-3 smaller than in the standard power spectrum,
yielding 10 times more cumulative signal-to-noise at $z=0$.
\end{abstract}


\section{Introduction}
The clustering of galaxies on extragalactic scales is a fundamental
cosmological observable.  But most cosmological analysis of galaxy
power spectra has been confined to the largest, linear scales, with
wavenumber $k \lesssim 0.2$\ihmpc\ \citep[\eg][]{pope, tegmark}.  This
is because various effects obscure cosmological information on
nonlinear scales: scale-dependent bias between the observable galaxies
and the theoretically more straightforward dark-matter distribution;
redshift distortions from peculiar velocities; and the fact that
currently, N-body simulations are necessary to model the matter power
spectrum accurately on nonlinear scales.

In principle, these effects can be modeled, but recently, another
effect has been investigated that further discourages prospectors
looking on translinear scales, 0.2\ihmpc\ $\lesssim k \lesssim$
0.8\ihmpc, for cosmological information.  Translinear scales are not
fully linear, but they are larger than the characteristic scales of
halos.  Na\"{\i}vely then, one expects the nonlinearities there to be
weak, but the power spectrum variance and covariance are surprisingly
large \citep{mw,szh,coorayhu}.  This makes the gain in cosmological
parameter Fisher information modest if an analysis of the large-scale
power spectrum is extended to include translinear scales
\citep{rh05,rh06,ns06,ns07,leepen,takahashi}.  At least in a halo
model \citep[\eg][]{cooraysheth} of large-scale structure, this effect
comes from cosmic variance in the halo population: a chance
preponderance of large halos in a survey increases the power spectrum
disproportionately on translinear scales \citep{ns06}.

\begin{figure}
  \begin{center}
    \includegraphics[scale=0.40]{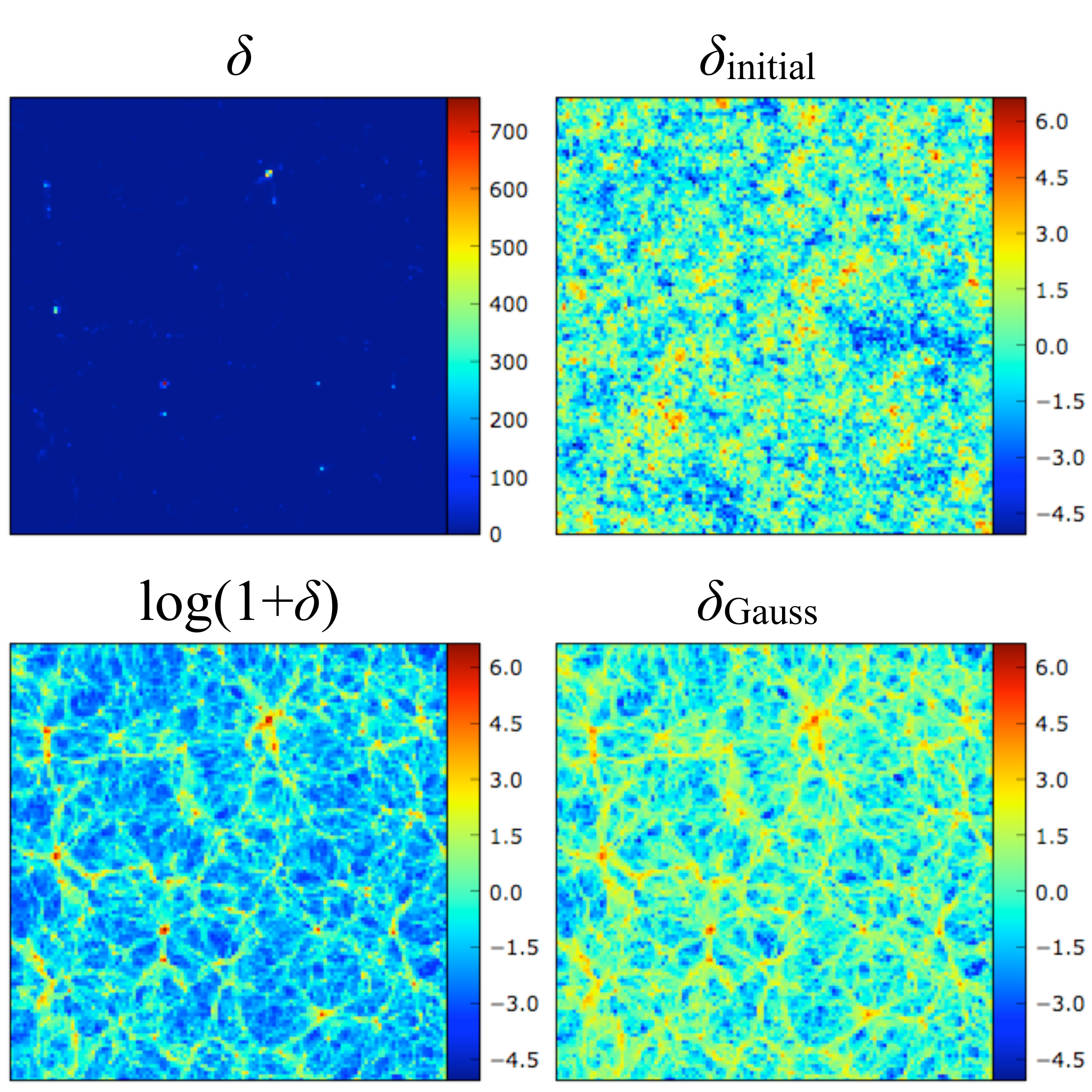}
  \end{center}

  \caption{Dark matter density in a slice 2\hmpc\ thick and
    250\hmpc\ on a side in the Millennium simulation.  $\delta$ is the
    overdensity at $z=0$.  $\delta_{\rm init}$ is at $z=127$, scaled
    to have the same minimum as $\log(1+\delta)$, the natural
    logarithm of the density at $z=0$. The Gaussianized density \dg,
    at $z=0$, preserves the ranking of cell densities but imposes a
    Gaussian PDF with $\Var(\delta_{\rm Gauss})=\Var[\log(1+\delta)]$.
    Particular structures at $z=0$ and $z=127$ disagree because of
    drift as the simulation progresses.  See {\tt
      http://skysrv.pha.jhu.edu/\til neyrinck/sonifylss.html} for
    sounds representing these panels.  The similarity in the sounds
    (except $\delta$'s) suggests a similarity in their power spectra.}
  \label{fig:showdens}
\end{figure}

\begin{figure}
  \begin{center}
    \includegraphics[scale=0.40]{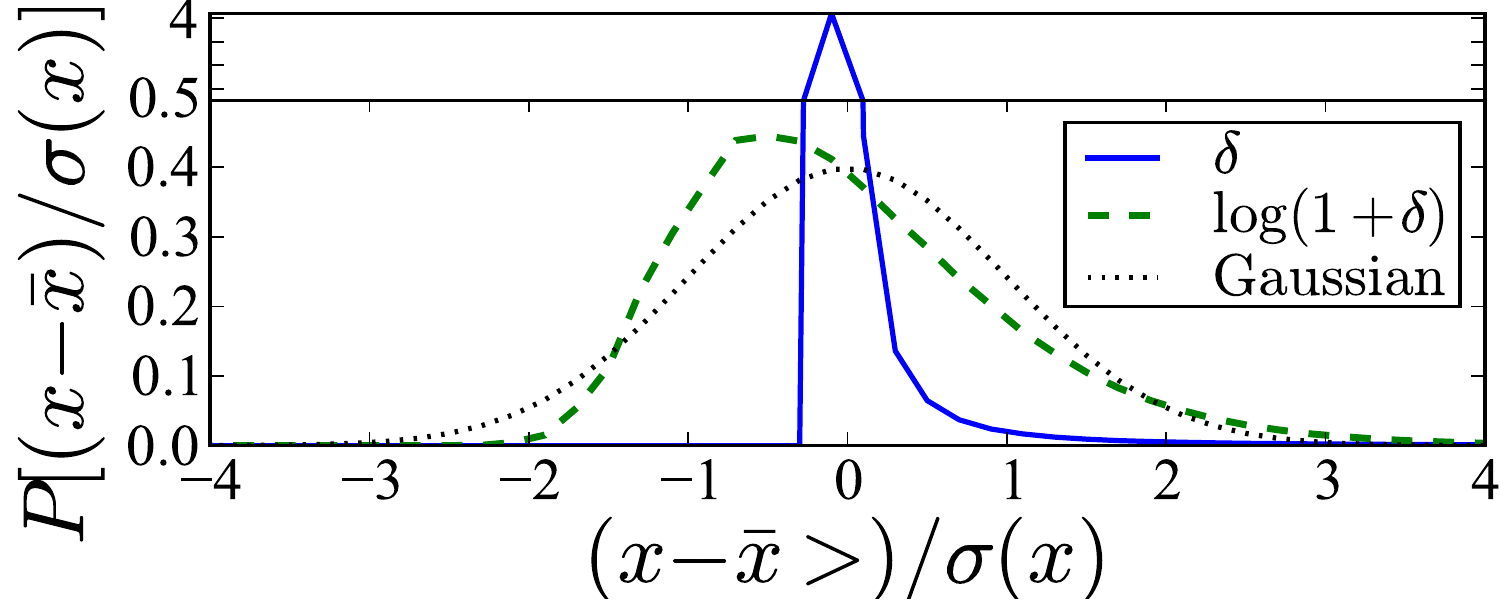}
  \end{center}  
  \caption{PDF's of $\delta$ and $\log(1+\delta)$ measured in
    $2$-\hmpc\ cells in the Millennium simulation at $z=0$.  Each
    distribution has been calibrated to have standard deviation 1.}
  \label{fig:lognormality}
\end{figure}

Figure \ref{fig:showdens} shows a slice of the Millennium Simulation
density field \citep[MS;][]{mill}.  The bottom panels show
$\log(1+\delta)$, as well as a Gaussianized \dg, in which the ranking
of cell densities is preserved, but they are mapped to a Gaussian
probability density function (PDF).  (`log' denotes the natural
logarithm.)  The bottom panels show more structure than the $\delta$
panel, which is why simulation visualizations often use logarithmic
color tables.  Figure \ref{fig:lognormality} shows the high
non-Gaussianity of the $\delta$ PDF compared to the $\log(1+\delta)$
PDF.

All that is visible in the $\delta$ panel is a handful of discrete
peaks (halos).  This suggests why the standard power spectrum, \pd, is
so sensitive to fluctuations in the halo population, and also suggests
why the halo model successfully describes \pd.  The other panels
contain obvious filamentary structure, which we suspect the halo model
would have more difficulty describing.

There are several reasons to use the log transform, despite the slight
non-lognormality of the density PDF \citep{colombi}.  The lognormal
distribution is simple; its use in galaxy number density PDF's dates
back to \citet{hubble}.  The log transform is easily reversible, and
thus preserves pixel-by-pixel information.  \citet{colesjones} showed
that a lognormal density PDF emerges if peculiar velocities are
assumed to grow according to linear theory, and showed that a
lognormal density PDF can explain many features observed in our
Universe.

In Schr\"{o}dinger perturbation theory \citep[SPT;][]{spt},
$A=\log(1+\delta)/2$ is a natural density variable.  In determining
the variance of $\delta$, tree-level SPT in $A$ captures most of the
higher-order loop corrections in standard perturbation theory.  This
suggests that the power spectrum of the log-mapped density could pull
in information from higher-order statistics of $\delta$.

Another motivation for using the log transform is to make the power
spectrum more suited to describe the density field.  The power
spectrum contains all the cosmological information in the Gaussian
initial conditions; all higher moments are zero.  The density field is
statistically invariant under translations and rotations at all
epochs.  Initially, there is also a symmetry between underdense and
overdense regions, related to the initial Gaussianity of the density
PDF.  At late times, this symmetry is broken, and high overdensities
receive perhaps undue weight in measurement of the power spectrum.  In
this paper, we test the hypothesis that restoring the Gaussianity of
the density PDF restores some information to the power spectrum.

\section{Power spectrum nonlinearities}
\label{sec:results}
We measure nonlinearities in power spectra of the log-density (\plog)
and the Gaussianized overdensity (\pg) using the dark-matter density
field from the 500 \hmpc\ \LCDM\ Millennium simulation (MS).  We use
publicly available time snapshots of the density measured with a
nearest-grid-point (NGP) method on a 256$^3$ grid.  On this mesh, the
shot noise is negligible; there is a mean of 600 and minimum of 6
particles per grid cell.

\begin{figure}
  \begin{center}
    \includegraphics[scale=0.40]{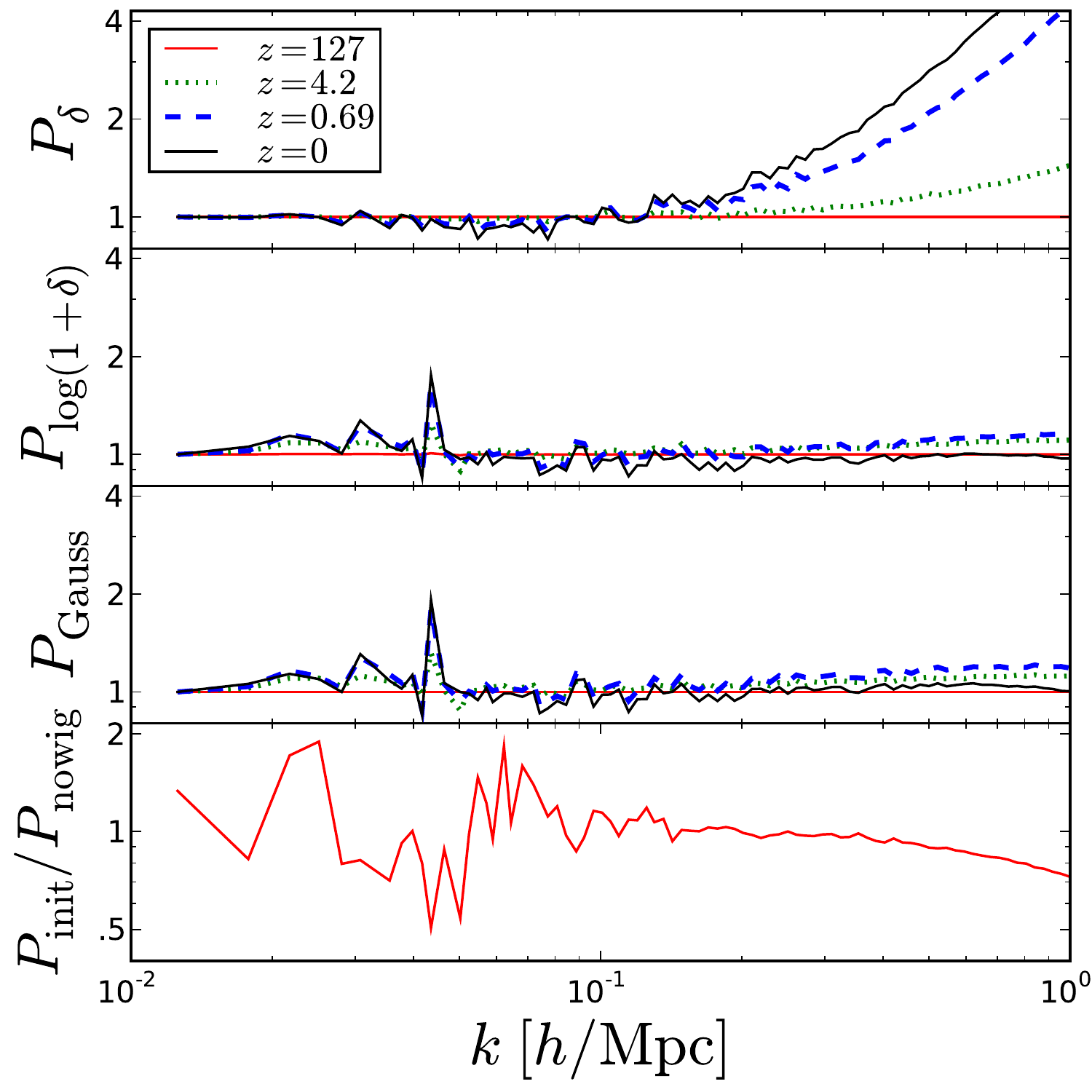}
  \end{center}  
  \caption{The top 3 panels show the ratios of \pd, \plog, and \pg\ to
    the initial power spectrum at various redshifts of the Millennium
    simulation.  We apply a multiplicative factor to each curve to
    line up the power spectra in the lowest-$k$ bin.  The bottom panel
    shows the level of fluctuation in the initial power spectrum using
    the same bins, relative to a no-wiggle power spectrum.}
  \label{fig:powers}
\end{figure}

Figure \ref{fig:powers} compares the distortion gravitational
evolution imparts to various power spectra in the MS.  We show the
ratio of each power spectrum to the initial-conditions ($z=127$) power
spectrum, dividing by factors to line up the power spectra in the
lowest-$k$ bin.  Also shown are the fluctuations of the initial power
spectrum with the same binning, dividing by the no-wiggle power
spectrum of \citet{ehu}.  We do not correct for the NGP pixel window
function; for example, this causes the small-scale downturn in the
bottom panel.  We cut the plot slightly before the Nyquist wavenumber
($k=1.6$\ihmpc), but effects from the NGP assignment may persist at
the smallest scales plotted.  However, these effects should be mild
for the ratios of two similar power spectra, i.e.\ in the \plog\ and
\pg\ panels.

Both \plog\ and \pg\ experience minimal nonlinearities compared to the
standard power spectrum \pd, to the extent that we conjecture that
they trace the linear power spectrum itself well into the nonlinear
regime.  Possibly, a different transformation than a logarithm could
give a better estimator of the linear power spectrum, but the
similarity of \plog\ and \pg\ to each other suggests that \plog\ is
close to optimal already.

\begin{figure}
  \begin{center}
    \includegraphics[scale=0.40]{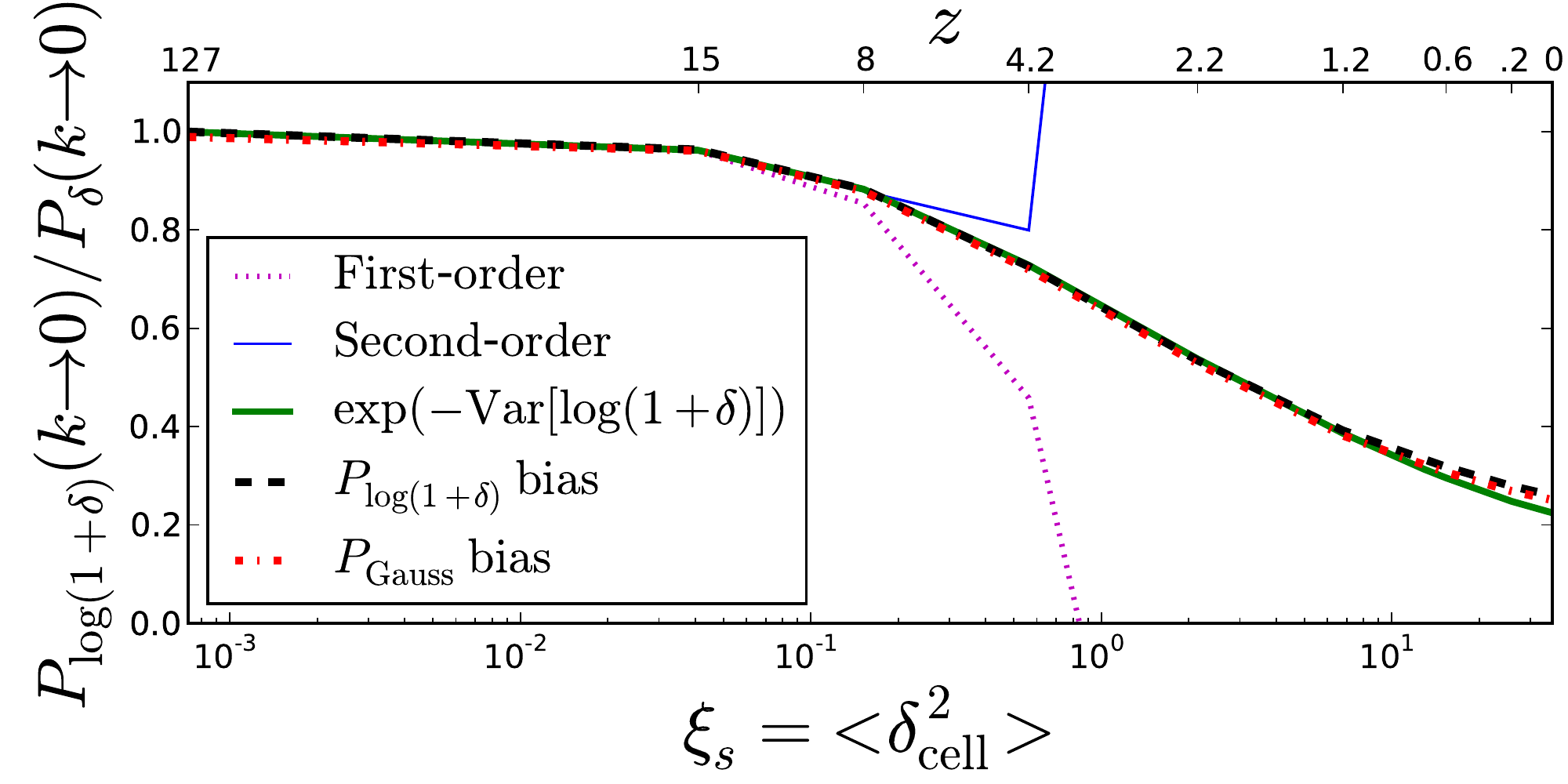}
  \end{center}  
  \caption{The large-scale biases \plog/\pd\ and \pg/\pd\ in the
    Millennium simulation sampled on a $256^3$ grid at various
    redshifts, measured using the lowest-$k$ bin.  Also shown are a
    few analytic approximations, as given in the Appendix.  The result
    in Eq.\ (\ref{eqn:bias}) traces each bias almost perfectly through
    $\avg{\delta_{\rm cell}^2} \approx 7$.}
  \label{fig:biases}
\end{figure}

As structure develops, a multiplicative bias grows between \plog\ and
\pd\ on large scales, shown in Figure \ref{fig:biases} for several
redshifts of the MS.  This bias depends on the variance of cell
densities, and thus on the cell size used to sample the density field.

In the Appendix, we estimate the large-scale bias between \plog\ and
\pd\ analytically, both perturbatively (assuming small fluctuations)
and non-perturbatively (assuming a lognormal density PDF).  Figure
\ref{fig:biases} shows the performance of these bias approximations.
The non-perturbative result,
\begin{equation}
  \lim_{k\to 0}\frac{P_{\log(1+\delta)}(k)}{P_{\delta}(k)} = e^{-\Var[\log(1+\delta_{\rm cell})]},
  \label{eqn:bias}
\end{equation}
works almost perfectly through $\avg{\delta_{\rm cell}^2} \approx 7$.
For the MS at this cell size, this occurs at
$z=1.2$.  At lower $z$, presumably, the density becomes insufficiently
lognormal for the calculation to work perfectly.  The perturbative
results, given in the Appendix, work for $\avg{\delta_{\rm cell}^2}\ll
1$, as expected.  We use perturbation theory
\citep[\eg][]{bernardeau} on a linear power spectrum from
\camb\ \citep{camb} to calculate the cumulants in the Appendix.  The
radius of the top-hat filter we used is that of a sphere with the
volume of a grid cell.  The bias is almost identical for \pg, if
$\Var(\delta_{\rm Gauss})$ is set to $\Var[\log(1+\delta)]$.

\section{Information content}
We compare the information contents of \plog\ and \pd\ using the
signal-to-noise \citep[S/N;][]{takahashi}, the Fisher information
\citep{fisher, tth} about the mean of a power spectrum itself.  The
S/N in a power spectrum over a range of bins $\R$ is defined as
\begin{equation}
  F(\R) = \sum_{i,j\in \R} (\bssC_\R^{-1})_{ij}.
  \label{inforange}
\end{equation}
$\bssC_\R$ is the covariance matrix of the power spectrum in bins,
$C_{ij}=\avg{(P_i-\bar{P_i})(P_j-\bar{P_j})}/(\bar{P_i}\bar{P_j})$,
not to be confused with the cumulants in the Appendix.  Fisher
information usually references a set of cosmological parameters, but
an analysis of how \plog\ varies with cosmological parameters is
beyond the scope of this paper.

To measure the covariance matrices, we subject the MS to 248 different
large-wavelength sinusoidal weightings that estimate the covariance in
a single cubic simulation \citep{hrs}.  This includes higher-order
weightings than the 52 they recommend, reducing the noise somewhat,
but increasing the minimum usable $k\in\R$.  For $\bar{P_i}$, we use
the power spectrum of the unweighted density.  For \plog, we apply the
weightings to $\log(1+\delta)-\avg{\log(1+\delta)}$, subtracting off
the mean because the weightings can cause a spike in the power
spectrum if the mean is nonzero.

\begin{figure}
  \begin{center}
    \includegraphics[scale=0.40]{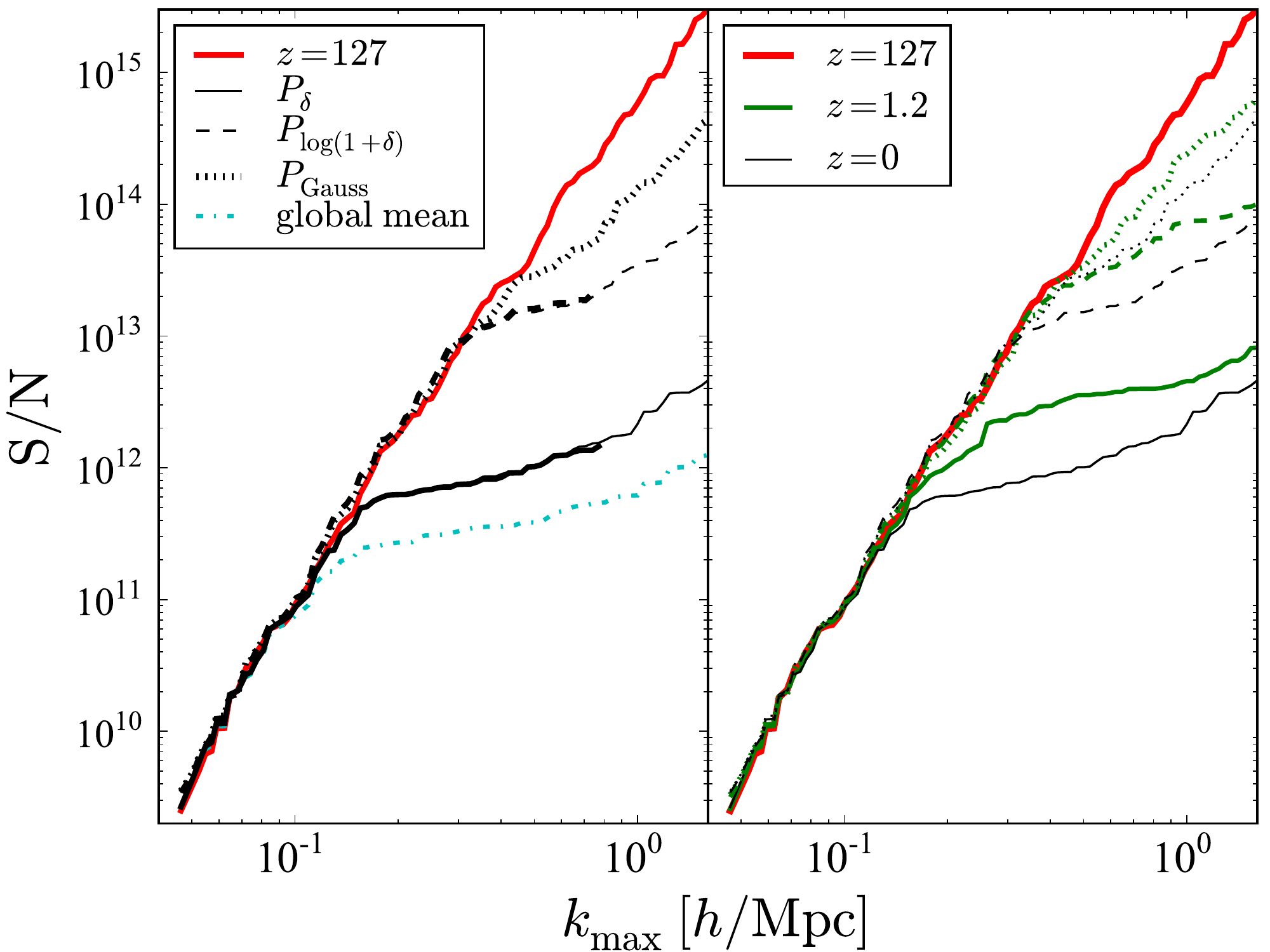}
  \end{center}  
  \caption{A comparison of the signal-to-noise for different power
    spectra at $z=0$, 1.2 and 127.  Solid, dashed, and dotted curves
    show S/N in \pd, \plog, and \pg.  The bold black curves use a
    degraded $128^3$ (from $256^3$) grid.  At left, S/N are at $z=0$;
    the right panel also shows S/N at $z=1.2$.  The \pdglob\ curve
    shows S/N for \pd\ if the global instead of local mean
    density is used to estimate $\delta$ for each weighting.
}
  \label{fig:info}
\end{figure}

Figure \ref{fig:info} shows S/N for various power spectra.  The
smallest $k\in\R$ is fixed at the smallest $k$ not directly affected
by weightings, and we vary the maximum $k_{\rm max}$.  On linear
scales, S/N $\propto k_{\rm max}^3$, proportional to the number of
modes.  We plot the S/N to the Nyquist wavenumber, because results for
a downgraded $128^3$ grid were nearly identical to the $256^3$ results
up to the $128^3$ Nyquist wavenumber.  Still, features at $k_{\rm
  max}\approx 1$\ihmpc\ should be interpreted cautiously.  For
\pdglob, the mean density $\bar{\rho}$ used in
$\delta=\rho/\bar{\rho}-1$ is the global mean of the simulation, while
for $P_{\delta}$, the `local' mean density is used,
i.e.\ $\avg{\rho(\br)w(\br)^2}$, where $w(\br)$ is one of the 248
weighting functions.  Using the local mean actually boosts the S/N
somewhat in \pd, because generally, denser regions have higher
translinear power.

The S/N in \plog\ continues to grow with $k_{\rm max}^3$ for an extra
factor of 2-3 beyond where the S/N in \pd\ turns over, giving a factor
of $\sim 10$ S/N increase on small scales.  The S/N in \pg\ reaches a
factor of 2 even higher.  The global vs.\ local mean-density issue
does not apply to \pg\ and \plog.

\section{Conclusion}
We find that the power spectrum of the log of the matter density,
\plog, and the power spectrum of $\delta$ after Gaussianizing its PDF,
\pg, suffer dramatically smaller nonlinearities than the standard
matter power spectrum \pd\ on translinear scales at all redshifts
tested.  This is true for both the mean of the power spectrum and its
covariance, as we measure from the high-resolution Millennium
simulation (MS).  Not only do \plog\ and \pg\ seem to trace the linear
power spectrum to nearly $k=1$\ihmpc, but they respectively contain
factors of 10 and 20 higher signal-to-noise than \pd\ at $z=0$.

To constrain cosmology, it is necessary to study how \plog\ varies
with cosmological parameters.  Much current large-scale-structure
research is focused on baryon acoustic oscillations (BAO's), so a
major issue to investigate is the degree of BAO attenuation in \plog.
The MS by itself is too small to provide an adequate answer to this
question.  BAO detection would likely benefit from the lack of
nonlinearities in \plog\ on BAO scales, although the BAO's in
\plog\ might experience some smearing as in \pd.  These results are
also encouraging for non-BAO large-scale-structure cosmology, since
they imply a significantly larger range of scales over which the
linear power spectrum can be accessed.

It is also necessary to investigate the effects of redshift space
distortions, shot noise, and galaxy bias on \plog.  Shot noise could
be a significant issue in realistic surveys because accurately
constructing the $\log(1+\delta)$ and \dg\ fields requires knowledge
of the density even in voids.  It could be that \plog\ has low
sensitivity to the accuracy of densities estimated in voids, but it is
likely that adaptive density kernels or tessellation methods
\citep[\eg][]{vdws} will be needed to achieve substantial information
gains with \plog\ over \pd.  The issue of galaxy bias could actually
be a positive feature of \plog\ even on linear scales, since at least
in the limit of large $\delta\approx \delta+1$, \plog\ is insensitive
to linear galaxy bias.  Although \pg\ outperforms \plog\ in our
idealized simulation, it remains to be seen what the optimal transform
is in the face of shot noise and other observational issues.

It is reassuring to find that at least in principle, much information
that gravity seemed to have unfairly stripped from the matter power
spectrum on translinear scales is retrievable with a simple
transformation on the density field.  A log transform could also prove
useful for other statistics, such as estimators of primordial
non-Gaussianity from large-scale structure.

\acknowledgments We thank Andrew Hamilton for useful discussions.
Some results used the CosmoPy package
(\url{http://www.ifa.hawaii.edu/cosmopy}). The Millennium Simulation
databases used in this paper and the web application providing online
access to them were constructed as part of the activities of the
German Astrophysical Virtual Observatory. The idea for this paper
germinated at the stimulating Aspen Center for Physics.  MN and AS are
grateful for support from the W.M.\ Keck and the Gordon and Betty
Moore Foundations, and IS from NASA grant NNG06GE71G, NSF grant
AMS04-0434413, and the Pol\'anyi Program of the Hungarian National
Office for Research and Technology (NKTH).

\bibliographystyle{hapj}
\bibliography{refs}

\appendix
\section{Analytic estimates of the log-density power spectrum bias}
The power spectrum $P_A$ of the log-density $A =\log(1+\delta)$ [for
simplicity, $A$ is twice the quantity defined by \citet{spt}] is
generally biased on large scales relative to the standard $P_\delta$.
We can calculate the moments of the $A$ field with brute-force
perturbation theory for small $\delta$, and express them in terms of
the connected moments of $\delta$.

Since $A$ is not a zero-mean field, we start with the mean for
completeness.  Expanding the logarithm:

\begin{equation}
\avg{A} = \avg{\delta-\frac{\delta^2}{2}+\frac{\delta^3}{3}-\frac{\delta^4}{4}\pm\ldots}= \frac{\xi_s}{2}+\frac{S_3\xi_s^2}{3}+\frac{3\xi_s^2}{4} + 
{\cal O}(\xi_s^3),
\end{equation}
where $\xi_s$ is the average of the correlation function within grid
cells (i.e.\ the variance of cell densities). The main
subtlety of the calculation is to take into account the connectedness
of the moments \citep[\eg][for detailed explanation]{Szapudi2009}.  The skewness $S_3$ and the cumulants
$C_{N,M}$ (below) are defined by $S_3=\frac{\avg{\delta^3}_{\rm
    c}}{\avg{\delta^2}^2}$ and
$C_{N,M}=\frac{\avg{\delta_1^N\delta_2^M}_{\rm
    c}}{\avg{\delta_1\delta_2}\avg{\delta^2}^{M+N-2}}$.

The calculation of $\avg{A_1A_2}$ is analogous to the above.
Expanding the logarithm, and expressing the results in terms of
connected moments:

\begin{equation}
\avg{A_1A_2}=\frac{\xi_s^2}{4}+\xi_\ell \left[1+\xi_s \left(2-C_{1,2}\right)+\xi_s^2 \left(7-2 S_3-4 C_{1,2}+\frac{2 C_{1,3}}{3}+\frac{C_{2,2}}{4}\right)\right]+{\cal O}(\xi_s^3).
\end{equation}
The DC term $\xi_s^2/4$ affects the two-point correlation function,
but not the power spectrum.  The large-scale power spectrum bias is
the coefficient of $\xi_\ell$, the large-scale correlation function
subtracting off contributions from $\xi_s$.

For the non-perturbative calculation, we introduce $D = A-\avg{A}$.
$\avg{\delta} = 0$, so $\avg{e^{D+\avg{A}}}=1$.  Using the connected
moment theorem \citep{generatingfunctionology,Szalay88,szapudiszalay},
and assuming $D$ to be Gaussian, making all of its
higher-than-second-order connected moments vanish, we obtain
\begin{equation}
1+\xi_\ell = \avg{e^{D_1}e^{D_2}} = e^{\avg{D_1 D_2}+\avg{D^2}}.
\end{equation}
Using $\exp{\avg{D_1 D_2}} \simeq 1+\avg{D_1 D_2}$, and Fourier
transforming, yields the non-perturbative shift of $\exp{\avg{D^2}}$ in
Eq.\ (\ref{eqn:bias}).

\end{document}